\documentclass[12pt]{article}
\usepackage{makeidx}
\usepackage{amsfonts}
\usepackage{amssymb}
\usepackage{amsmath}
\usepackage[numbers,sort&compress]{natbib}
\usepackage[colorlinks,citecolor=blue,urlcolor=blue]{hyperref}
\usepackage{graphicx,indentfirst,tabularx}
\usepackage[bottom]{footmisc}
\usepackage{subcaption}

\newtheorem{theorem}{Theorem}

\newtheorem{Algorithm}[theorem]{Algorithm}

\input{tcilatex}
\begin{document}

\title{Bayesian Modeling of Gibbs Point Processes via Basis Function Expansions}

\author{Christopher Hassett\footnote{\baselineskip=10pt Veterans United, Columbia, Missouri 65203, USA, email: hassett.chris.m@gmail.com}, Athanasios C. Micheas\footnote{\baselineskip=10pt Department of Statistics and Data Science, University of Missouri, Columbia, Missouri 65211, USA, email: micheasa@missouri.edu}, Scott H. Holan\footnote{\baselineskip=10pt Department of Statistics and Data Science, University of Missouri, Columbia, Missouri 65211, USA, email: holans@missouri.edu},
and\\ Stamatis Dostoglou\footnote{\baselineskip=10pt Department of Mathematics, University of Missouri, Columbia, Missouri 65211, USA, email: dostoglous@missouri.edu}}
\maketitle

\begin{abstract}
We present a hierarchical Bayesian framework for non-homogeneous pairwise interaction Gibbs point process models, where the global and local effect functions are modeled via basis function expansions. We further propose a testing procedure in order to assess complete spatial randomness. The proposed methodology is exemplified through two real benchmark data examples involving water striders and forest fires.
\end{abstract}


\section{Introduction}
	\label{CH_Intro}


A point process (PP) is a random collection of events (points), where the number of events as well as their locations are random. The observation of such a random set is known as a point pattern. For modeling, computation methods and applications of PP models see \cite{Micheas2014}, \cite{zhou2015spatio},\cite{Micheas2019}, \cite%
{ChenMicheasHolan2020}, \cite{Baddeley2022}, \cite{WuMicheas2022}, \cite{chen2022hierarchical},
\cite{Baetal2023}, \cite{WuMicheas2024}, \cite{Micheas2025}, and \cite{wang2026inference} and the references therein.

Poisson and Gibbs PP models in particular, are defined using an intensity function and a conditional intensity function known as the Papangelou conditional intensity \cite{papangelou1974conditionalIntensity}, respectively, since these functions uniquely determine the probability distribution of the random point pattern. \cite{zammit-mangion_etal2012ppmodellingofAWD} used a basis function expansion to model the intensity of a Poisson point process (PPP). \cite{Jalilian2019} proposed an orthogonal series expansion of the pairwise correlation function. More recently, \cite{Baetal2023} used basis expansion in Gibbs point processes models, under the somewhat restrictive assumption of stationarity. In this paper, we generalize these existing methods to the general case of the pairwise interaction model Gibbs PP, that includes the PPP as a special case.

Specifically, we assume a model with a global effect term (which governs the locations of the events in space) and a pairwise interaction term (which describes how the events interact with each other), and taking basis function expansions of the functions in these terms we obtain a model that is highly flexible.  We illustrate the use of the Bernstein polynomial basis functions (\cite{tan2016bernstein_polynomials, curtis2011}) and propose specific monotonicity constraints on the basis coefficients we require in our modeling formulation. The coefficients of the basis functions are estimated using Liang's double Metropolis-Hastings algorithm \cite{liang2010doubleMH}.  More importantly, the methods proposed lend to the creation of a procedure that can be used to assess complete spatial randomness, which is another contribution of this work.

The paper is organized as follows.  In Section~\ref{SectionPoissonAndGibbsViaBasisExpansions}, we briefly discuss Poisson and Gibbs point processes along with our approach to modeling the global and local effect functions via basis expansions.  Section~\ref{SectionHBFramework} presents the Bayesian framework, the proposed model, and the sampling algorithm for sampling from the posterior distribution.  Furthermore, we present a method of assessing whether a Poisson (independent events) or a Gibbs (interacting events) process is more appropriate for a given point pattern. Section~\ref{SectionApplications}, illustrates the models for real data involving water striders and forest fires.  Concluding remarks are given in Section~\ref{SectionConclusionAndFutureWork}.

\section{Poisson and Gibbs Point Processes via Basis Expansions}
\label{SectionPoissonAndGibbsViaBasisExpansions}


Assume that the events are observed over a bounded window (domain) $W \subset \mathbb{R}^{2}$.  We can specify a point process through its density or conditional intensity.  If $f_{n}(\cdot)$ is the density of the point process, then the Papangelou conditional intensity \cite{papangelou1974conditionalIntensity} is defined as $\lambda(\mathbf{x} | \varphi_{n}) = f_{n}(\varphi_{n} \cup {\mathbf{x}}) / f_{n}(\varphi_{n}),$ where $\mathbf{x}$ is a typical event and $\varphi_{n}= \{ \mathbf{x}_{1}, \dots ,\mathbf{x}_{n} \}$ is the observed point pattern.
The Papangelou conditional intensity gives the conditional intensity of a point at location $\mathbf{x}$ given the point pattern $\varphi_{n}$.  Note that if $\mathbf{x}$ is in $\varphi_{n},$ then we define $\lambda(\mathbf{x} | \varphi_{n}) = 0.$

The Gibbs Point Process (PP) is defined via the main (global) and the interaction (local) effect functions $h(\mathbf{x}),$ $\mathbf{x} \in \mathbb{R}^{2}$ and $g(r),$ $r > 0,$ respectively.  In particular, the density of $n$ events is given by
\begin{equation}
f_{n}(\mathbf{x}_{1}, \dots ,\mathbf{x}_{n} ) = \alpha \, \text{exp} \left \{ -\sum_{i=1}^{n} h(\mathbf{x}_{i}) - \sum_{\substack{i,j = 1 \\ i < j}}^{n} g(||\mathbf{x}_{i} - \mathbf{x}_{j}||) \right \}, \label{eqf}
\end{equation}
where $\alpha$ is the normalizing constant, so that the Papangelou is given by
\begin{equation}
\lambda(\mathbf{x} | \mathbf{x}_{1}, \dots ,\mathbf{x}_{n} ) = \text{exp}\left \{ -h(\mathbf{x}) - \sum_{i=1}^{n} g(||\mathbf{x} - \mathbf{x}_{i}||) \right \}. \label{eqpap}
\end{equation}
Now, when $g(r) = 0,$ $\forall r > 0,$ the points are independent of each other, which yields the Poisson PP for independent events.

\subsection{Basis Function Expansions}
Mathematically, in order to express an expansion of a real-valued function $f,$ of one or two dimensions, we write \begin{equation}
f(x) = \sum_{i=1}^{\infty} \alpha_{i} \phi_{i}(x),\text{ and } f(x,y) = \sum_{i=1}^{\infty} \sum_{j=1}^{\infty} \alpha_{ij} \phi_{ij}(x,y),
\end{equation}
where in the one dimensional case $\phi_{i}(x)$ is the $i$th basis function and $\alpha_{i}$ is its associated coefficient, and for two dimensions $i$ corresponds to the $x$ dimension and $j$ corresponds to the $y$ dimension and $\phi_{ij}$ the basis functions with $\alpha_{ij}$ its coefficients. Consequently, we can approximate a variety of unknown functions with a finite number of basis functions by estimating their basis coefficients.

While many different basis functions could be used, we choose to use the Bernstein polynomials \cite{curtis2011}.  This choice is made due to the simplicity of adding constraints on the coefficients to enforce a desired behavior for $f(\cdot)$ \cite{curtis2011,tan2016bernstein_polynomials}.  Details of the Bernstein polynomials and the required constraints are provided in Section~\ref{SectionExParams_BernsteinPolynomial}.

\subsection{Potential Issues With Modeling Both Global and Interaction Effects}
\label{Section_Identifiability}
When splitting the model into a global effect and an interaction effect, one needs to be aware of the possibility of identifiability issues.  \cite{baddeley2000non} illustrated that the global effect and local effect can be difficult to distinguish, and one remedy is to make assumptions about the smoothness of the global effect.  More discussion on the potential for identifiability problems is provided in \cite{diggle2007second}.  As a way to deal with this issue, they suggest constructing the local effect on a smaller scale than the global effect.

Following \cite{baddeley2000non} and \cite{diggle2007second}, we define the global effect over the entire domain, and the number of global effect basis functions will be kept small, so that the global effect remains a smooth function. The basis functions for the interaction effect will be scaled and truncated, so that there is no interaction between points that are too far apart.  We then include enough basis functions to allow for a variety of interaction functions.

This flexible approach can handle a variety of choices for the functions $h: \mathbb{R}^{2} \rightarrow \mathbb{R}_{0}^{+}$ and $g: \mathbb{R}_{0}^{+} \rightarrow \mathbb{R}_{0}^{+}$, where $\mathbb{R}_{0}^{+}$ denotes the positive real numbers augmented with zero.  The approximate basis function expansions of $h(\mathbf{x})$ and $g(r)$ are given by
\begin{equation}
h(\mathbf{x}) \approx \sum_{i=1}^{i_{1}} \sum_{j=1}^{j_{1}} \alpha_{ij} \phi_{ij}(\mathbf{x}), \label{eqh}
\end{equation}
and
\begin{equation}
g(r) \approx \sum_{k=0}^{k_{1}} c_{k} b_{k}(r), \label{eqg}
\end{equation}
where $\alpha_{ij}$ and $c_{k}$ are the coefficients and $\phi_{ij}$ and $b_{k}$ are the associated basis functions for the expansion of the global and local effect functions, respectively.  The index $i_{1}$ represents the largest index for the global effect basis functions in the $x$ direction, whereas, the index $j_{1}$ represents the largest index for the global effect basis functions in the $y$ direction.  The index $k_{1}$ represents the largest index for the interaction effect basis functions.
For what follows, we set $i_{1} = j_{1}.$

The basis functions and their coefficients can be vectorized for mathematical convenience as follows; we define
\begin{align*}
\boldsymbol{\phi}(\mathbf{x}) = \left( \phi_{1}(\mathbf{x}), \ldots, \phi_{L}(\mathbf{x}) \right)' &= \left( \phi_{1,1}(x, y), \phi_{1,2}(x, y), \ldots, \phi_{i_{1}, j_{1}}(x, y) \right)', \\
\boldsymbol{\alpha} = \left( \alpha_{1}, \ldots, \alpha_{L} \right)' &= \left( \alpha_{1,1}, \alpha_{1,2}, \ldots, \alpha_{i_{1}, j_{1}} \right)', \\
\mathbf{b}(r) = \left( b_{1}(r), \ldots, b_{M}(r) \right)' &= \left( b_{1}(r), \ldots, b_{k_{1}}(r) \right)', \\
\end{align*}
and $\mathbf{c}=\left( c_{1}, \ldots, c_{M} \right)' = \left( c_{1}, \ldots, c_{k_{1}} \right)',$
with $L = i_{1} j_{1}$ and $M = k_{1}$ being the number of basis functions.  Consequently, (\ref{eqh})~and~(\ref{eqg}) can be written as $h(\mathbf{x}) \approx \boldsymbol{\phi}(\mathbf{x})' \boldsymbol{\alpha}$ and $g(r) \approx \mathbf{b}(r)' \mathbf{c},$  and as a result, the Papangelou from (\ref{eqpap}) is approximated by
\begin{align*}
\lambda(\mathbf{x} | \varphi_{n}) &\approx \text{exp} \left \{ -\boldsymbol{\phi}(\mathbf{x})' \boldsymbol{\alpha} - \sum_{i=1}^{n} g(||\mathbf{x} - \mathbf{x}_{i}||) \right \} \\
&= \text{exp} \left[ -\boldsymbol{\phi}(\mathbf{x})' \boldsymbol{\alpha} - \left \{ \sum_{k=1}^{n} \mathbf{b}(||\mathbf{x} - \mathbf{x}_{k}||) \right \}' \mathbf{c} \right].
\end{align*}
Finally, the general model for Equation~(\ref{eqf}) can be written as
\begin{align*}
f_{n}(\varphi_{n}) 
&\propto \text{exp} \{ -\boldsymbol{\phi}(\mathbf{x}_{1})' \boldsymbol{\alpha} - \boldsymbol{\phi}(\mathbf{x}_{2})' \boldsymbol{\alpha} - \mathbf{b}(||\mathbf{x}_{2} - \mathbf{x}_{1}||)' \mathbf{c} \\
&- \boldsymbol{\phi}(\mathbf{x}_{3})' \boldsymbol{\alpha} - \mathbf{b}(||\mathbf{x}_{3} - \mathbf{x}_{1}||)' \mathbf{c} - \mathbf{b}(||\mathbf{x}_{3} - \mathbf{x}_{2}||)' \mathbf{c} - \cdots - \mathbf{b}(||\mathbf{x}_{n} - \mathbf{x}_{n-1}||)' \mathbf{c} \} \\
&\propto \text{exp} \left[ -\left\{ \sum_{i=1}^{n} \boldsymbol{\phi}(\mathbf{x}_{i}) \right \}' \boldsymbol{\alpha} - \left \{ \sum_{\substack{i < j \\ i,j \ge 1}}^{n} \mathbf{b}(||\mathbf{x}_{i} - \mathbf{x}_{j}||) \right \}' \mathbf{c} \right].
\end{align*}
Now, if $Z_{n}(\boldsymbol{\alpha}, \mathbf{c})$ denotes the normalizing constant, then the model is
\begin{align}
f_{n}(\varphi_{n}) &= \frac{1}{Z_{n}(\boldsymbol{\alpha}, \mathbf{c})} \text{exp}\left[-\left\{ \sum_{i=1}^{n} \boldsymbol{\phi}(\mathbf{x}_{i}) \right\}' \boldsymbol{\alpha} - \left\{ \sum_{\substack{i < j \\ i,j \ge 1}}^{n} \mathbf{b}(||\mathbf{x}_{i} - \mathbf{x}_{j}||) \right\}' \mathbf{c} \right] \label{eqf_expanded} \\
&= \frac{1}{Z_{n}(\boldsymbol{\alpha}, \mathbf{c})} f_{n}^{kern}(\varphi_{n}, \boldsymbol{\alpha}, \mathbf{c}), \nonumber
\end{align}
with $$Z_{n}(\boldsymbol{\alpha}, \mathbf{c}) = \int_{\mathbb{R}^{2n}} \exp\{-\text{En}(\varphi_{n}; \boldsymbol{\alpha}, \mathbf{c})\} d\mathbf{x}_{1} \cdots d\mathbf{x}_{n},$$ where
\begin{equation*}
\text{En}(\varphi_{n}, \boldsymbol{\alpha}, \mathbf{c}) = \left\{ \sum_{i=1}^{n} \boldsymbol{\phi}(\mathbf{x}_{i}) \right\}' \boldsymbol{\alpha} + \left\{ \sum_{\substack{i < j \\ i,j \ge 1}}^{n} \mathbf{b}(||\mathbf{x}_{i} - \mathbf{x}_{j}||) \right\}' \mathbf{c}
\end{equation*}
denotes the energy function of the $n$ events and $f_{n}^{kern}(\varphi_{n}) = \exp \left\{ -\text{En}(\varphi_{n}; \boldsymbol{\alpha}, \mathbf{c}) \right\}.$


\subsection{Choice of Basis Functions}
\label{SectionExParams_BernsteinPolynomial}
Before describing the basis functions, we assume, without loss of generality, that the point pattern is observed over the window $W = [0,1]^{2}$, since we can always transform the points to be in $W$. We use Bernstein polynomial basis functions \cite{curtis2011,tan2016bernstein_polynomials} in order to describe both the global and interaction effects.  We modify the traditional Bernstein polynomials, so that the indices start at one and go to the number of basis functions, instead of starting at zero and going to the number of basis functions minus one.  That is, let $$\phi_{ij}(\mathbf{x}) = \binom{i_{1} - 1}{i - 1} \binom{j_{1} - 1}{j - 1} x^{i - 1} (1 - x)^{i_{1} - i} y^{j - 1} (1 - y)^{j_{1} - j} I(0 \le x \le 1) I(0 \le y \le 1),$$
where $\mathbf{x} = (x,y)'$, and
\begin{equation}
b_{k}(r) = \binom{k_{1} - 1}{k - 1} \left(\frac{r}{r_{max}}\right)^{k - 1} \left(1 - \frac{r}{r_{max}}\right)^{k_{1} - k} I\left(0 \le \frac{r}{r_{max}} \le 1\right), \label{eq_b}
\end{equation}
with $i_{1}, j_{1}, k_{1}$ as in Equations (\ref{eqh}) and (\ref{eqg}).

Given the values of $x$ and $y$, we choose $r \in \left[0, r_{max}\right]$, where $$r_{max} = \text{median}_{1 \le i < j \le n}\left\{\sqrt{(x_{i} - x_{j})^2 + (y_{i} - y_{j})^2}\right\},$$ is the median distance between the observed points.  As discussed in Section~\ref{Section_Identifiability}, if the model is not constructed carefully, there is a possibility for identifiability issues between the global effect and the local effect.  In (\ref{eq_b}), the scaling of $r$ by $r_{max}$ helps reduce the chance for identifiability problems, by limiting the distance at which there can be interaction between points.  Furthermore, $i_{1}$ and $j_{1}$ will be chosen to be small, so that the global effect is a smooth function.  Critically, for this choice of basis functions, the accuracy depends on the behavior of the functions we are trying to reconstruct.  As further illustrated below, depending on the function under consideration, we may require more basis functions.

Finally, we require the interaction function $g(r)$ to be nonnegative and non-increasing in $r$, since we expect the interaction effect between two events to diminish the farther apart the two points are from each other.  To obtain a non-increasing and nonnegative interaction function, we require that the coefficients $\mathbf{c}$ of the basis expansion for the interaction effect be such that $c_{k} \ge 0$, for all $k,$ and $c_{k} \ge c_{k+1}$, for $k = 1,\ldots,M$.  We refer to the work of \cite{curtis2011} and \cite{tan2016bernstein_polynomials} for more details on choosing appropriate constraints depending on the properties required for $g(r).$

\section{Bayesian Framework and Posterior Simulation}
\label{SectionHBFramework}


We consider a Bayesian approach to modeling $\boldsymbol{\alpha}$ and $\mathbf{c}$.  The joint distribution of the data and the parameters, $f_{n}(\varphi_{n}, \boldsymbol{\alpha}, \mathbf{c}, \mu_{\alpha}, \sigma_{\alpha}^{2}, \lambda_{c}),$ can be decomposed into conditional distributions as follows,
\begin{equation*}
f_{n}(\varphi_{n}, \boldsymbol{\alpha}, \mathbf{c}, \mu_{\alpha}, \sigma_{\alpha}^{2}, \lambda_{c}) = f_{n}(\varphi_{n} | \boldsymbol{\alpha}, \mathbf{c}) f(\boldsymbol{\alpha} | \mu_{\alpha}, \sigma_{\alpha}^{2}) f(\mathbf{c} | \lambda_{c}),
\end{equation*}
where $\mu_{\alpha}, \sigma_{\alpha}^{2}, \lambda_{c}$ denote hyper-parameters.

The data model, $f_{n}(\varphi_{n}),$ which is the density for the point pattern, is obtained from (\ref{eqf_expanded}), and is of the form $$f_{n}(\varphi_{n}) = \frac{1}{Z_{n}(\boldsymbol{\alpha}, \mathbf{c})} \text{exp}\left[-\left\{ \sum_{i=1}^{n} \boldsymbol{\phi}(\mathbf{x}_{i}) \right\}' \boldsymbol{\alpha} - \left\{ \sum_{\substack{i < j \\ i,j \ge 1}}^{n} \mathbf{b}(||\mathbf{x}_{i} - \mathbf{x}_{j}||) \right\}' \mathbf{c} \right].$$
We assume that the basis coefficients $\boldsymbol{\alpha}$ are such that $\alpha_{i} \sim \text{N}(\mu_{\alpha}, \sigma_{\alpha}^{2})$, $i = 1,\ldots,L,$ and $\mathbf{c}$ follows the pdf of the exponential order statistics (EOS) \cite{casella_berger_2002_statistical_inference} in descending order with common rate $\lambda_{c},$ so that the prior distributions on the basis coefficients are given by
\begin{equation*}
\pi(\alpha_{i}|\mu_{\alpha}, \sigma_{\alpha}^{2}) \propto (\sigma_{\alpha}^{2})^{-\frac{1}{2}}\text{exp} \left \{ -\frac{1}{2} (\alpha_{i} - \mu_{\alpha})^{2} / \sigma_{\alpha}^{2} \right \},
\end{equation*}
and
\begin{equation*}
\pi(\mathbf{c}|\lambda_{c}) = \begin{cases}
n! f(c_{1}|\lambda_{c}) \cdots f(c_{M}|\lambda_{c}) & -\infty < c_{M} < \cdots < c_{1} < \infty \\
0 & \text{otherwise},
\end{cases}
\end{equation*}
for $i=1,\ldots,L,$ where each $c_{i}$ is distributed according to $f(c) = \frac{1}{\lambda_{c}} \exp(-c / \lambda_{c})$, for $c \ge 0$.  We specify the following parameter values, $\mu_{\alpha} = -10$, $\sigma_{\alpha}^{2} = 20,$ and $\lambda_{c} = 2.1.$  These choices for the hyper-parameters should be sufficiently uninformative, since from our extensive simulation studies (results not shown) a choice of $\boldsymbol{\alpha} = (-30, -30, \ldots, -30)$ would result in a very large intensity of points, whereas, choosing $\boldsymbol{\alpha} = (10, 10, \ldots, 10)$ would result in almost no points.
Based on the model distribution of (\ref{eqf}) and the aforementioned priors, the full posterior distribution of $\boldsymbol{\alpha}, \mathbf{c} | \cdot$ is given by
\begin{align}
\pi \left(\boldsymbol{\alpha}, \mathbf{c} | \cdot \right) &= \frac{1}{Z_{n}(\boldsymbol{\alpha}, \mathbf{c})} \text{exp}\left[ -\left \{ \sum_{i=1}^{n} \boldsymbol{\phi}(\mathbf{x}_{i}) \right \}' \boldsymbol{\alpha} - \left \{ \sum_{\substack{i < j \\ i,j \ge 1}}^{n} \mathbf{b}(||\mathbf{x}_{i} - \mathbf{x}_{j}||) \right \}' \mathbf{c} \right] \nonumber \\
&\times \prod_{i = 1}^{L} \left[(2\pi \sigma_{\alpha}^{2})^{-\frac{1}{2}} \exp\left\{ -\frac{1}{2} (\alpha_{i} - \mu_{\alpha})^{2} / \sigma_{\alpha}^{2} \right\}\right] n! \lambda_{c}^{M} \exp\left( -\lambda_{c} \sum_{j=1}^{M} c_{j} \right). \label{eqFullPosterior}
\end{align}

Consequently, the full conditional distribution of the $i$th global effect parameter is given by
\begin{align*}
\pi(\alpha_{i} | \varphi_{n}) &= \pi(\varphi_{n} | \boldsymbol{\alpha}, \mathbf{c}) \pi(\alpha_{i} | \mu_{i}, \sigma_{i}^{2}) \\
&\propto \frac{1}{Z_{n}(\boldsymbol{\alpha}, \mathbf{c})} \text{exp}\left[ -\left\{ \sum_{l=1}^{n} \phi_{i}(\mathbf{x}_{l}) \right\}' \boldsymbol{\alpha} \right] \exp\left\{ -\frac{1}{2} (\alpha_{i} - \mu_{i})^{2} / \sigma_{i}^{2} \right\},
\end{align*}
and full conditional distribution of the interaction parameters is given by
\begin{align*}
\pi(\mathbf{c} | \varphi_{n}) &= \pi(\varphi_{n} | \boldsymbol{\alpha}, \mathbf{c}) \pi(\mathbf{c} | \lambda_{c}) \\
&\propto \frac{1}{Z_{n}(\boldsymbol{\alpha}, \mathbf{c})} \text{exp}\left[ - \left\{ \sum_{\substack{i < j \\ i,j \ge 1}}^{n} \mathbf{b}(||\mathbf{x}_{i} - \mathbf{x}_{j}||) \right\}' \mathbf{c} \right] n! \lambda_{c}^{M} \exp\left( -\lambda_{c} \sum_{j=1}^{M} c_{j} \right).
\end{align*}


\subsection{Sampling Approach}
\label{SectionSampling}
In the full conditionals for $\alpha_{i}$ and $c_{j}$, $Z(\boldsymbol{\alpha}, \mathbf{c})$ cannot be calculated analytically.  Both \cite{moller_etal2006MCMCforIntractable} and \cite{liang2010doubleMH} provide methods for sampling $\alpha_{i}$ and $c_{j}$ using auxiliary variable techniques.  The former method is exact and computationally intensive, while the latter is an approximation that requires less computation, but still provides comparable accuracy.  To sample from the full conditionals we make use of the double Metropolis-Hastings algorithm proposed by \cite{liang2010doubleMH}.  The algorithm is as follows:

\begin{Algorithm}[Double M-H posterior sampler]
	\label{algDoubleMH}
	\quad \newline
	\textbf{Step 0:} Start with an initial state $\boldsymbol{\theta}^{(0)}=(\boldsymbol{\alpha}^{(0)},
\mathbf{c}^{(0)}$.  For all $k \ge 0$, assume that at time $k$ the Markov chain is at $\boldsymbol{\theta}^{(k)}=(\boldsymbol{\alpha}^{(k)},
\mathbf{c}^{(k)})$. \\
	\textbf{Step 1:}  Generate a proposed parameter vector $\boldsymbol{\alpha}^{(prop)} \sim P^{kern}\left( \cdot | \boldsymbol{\alpha}^{(k)}, \varphi_{n} \right)$, where the kernel (proposal) distribution is chosen to be symmetric, i.e., a $\text{N}(\boldsymbol{\alpha}^{(k)}, \Sigma_{\alpha, prop})$. Generate a proposed parameter vector $\mathbf{c}^{(prop)} \sim P^{kern}\left( \cdot | \mathbf{c}^{(k)}, \varphi_{n} \right)$, where the kernel distribution is once again $\text{N}(\mathbf{c}^{(k)}, \Sigma_{c, prop})$. \\
	\textbf{Step 2:}  Generate an auxiliary variable (point pattern) $\psi_{n} \sim f_{n}\left( \psi_{n} | \boldsymbol{\theta}^{(prop)} \right)$ using the Birth-Death algorithm with change moves.
	This simulation step corresponds to a first M-H step. \\
	\textbf{Step 3:}  The second M-H step occurs when we sample $\theta^{(k)},$ i.e., the M-H ratio is given by
	\begin{align*}
\small
	r\left( \boldsymbol{\theta}^{(prop)}, \boldsymbol{\theta}^{(k)}, \psi_{n}, | \varphi_{n} \right) = \frac{ \pi\left( \boldsymbol{\theta}^{(prop)} | \boldsymbol{\eta} \right) K\left( \boldsymbol{\theta}^{(k)} | \boldsymbol{\theta}^{(prop)}, \varphi_{n} \right) f_{n}\left( \psi_{n} | \boldsymbol{\theta}^{(k)} \right)  }{ \pi\left( \boldsymbol{\theta}^{(k)} | \boldsymbol{\eta} \right) K\left( \boldsymbol{\theta}^{(prop)} | \boldsymbol{\theta}^{(k)}, \varphi_{n} \right) f_{n}\left( \varphi_{n} | \boldsymbol{\theta}^{(k)} \right)  } \frac{f_{n}\left( \varphi_{n} | \boldsymbol{\theta}^{(prop)} \right)}{f_{n}\left( \psi_{n} | \boldsymbol{\theta}^{(prop)} \right)}
	\end{align*}
	and it does not depend on the normalizing constant $Z(\boldsymbol{\alpha}, \mathbf{c})$. Here $\boldsymbol{\eta}$ denotes all the parameters of the prior of $\boldsymbol{\theta}=(\boldsymbol{\alpha},
\mathbf{c})$.	
	
	\textbf{Step 4:}  Accept the proposed $\boldsymbol{\theta}^{(prop)}$ with probability $min\left\{ 1, r\left( \boldsymbol{\theta}^{(prop)}, \boldsymbol{\theta}^{(k)}, \psi | \varphi \right) \right\}$, and set $\boldsymbol{\theta}^{(k + 1)} = \boldsymbol{\theta}^{(prop)}$.  If the proposed value is not accepted, remain at the previous state, i.e., $\boldsymbol{\theta}^{(k + 1)} = \boldsymbol{\theta}^{(k)}$. Go to step 1.
\end{Algorithm}

\noindent In principle, $\boldsymbol{\alpha}$ and $\mathbf{c}$ could be sampled simultaneously (block sampling). However, this can lead to low acceptance rates and, therefore, we sample each $\alpha_{ij}$ and $c_{k}$ individually.


\subsection{Assessing Complete Spatial Randomness}


Now consider modeling inhomogeneous Poisson point processes (IPPPs), as a special case of the Gibbs model presented in Section~\ref{SectionPoissonAndGibbsViaBasisExpansions}, with the interaction function assumed to be zero.  We consider the same model and basis function expansion as above, using Bernstein polynomials. However, in order to assess independence (IPPP) versus dependence (Gibbs) we use a different prior on the coefficients of the expansion of the interaction effect.

Note that in order to have independent events, the interaction parameters must be $\mathbf{c}=\mathbf{0}$ in (\ref{eqf_expanded}) in which case the point pattern $\varphi_{n}$ is assumed to arise from an IPPP with intensity function $$f_{n}(\varphi_{n}) = \frac{1}{Z_{n}(\boldsymbol{\alpha})} \text{exp}\left[-\left\{ \sum_{i=1}^{n} \boldsymbol{\phi}(\mathbf{x}_{i}) \right\}' \boldsymbol{\alpha} \right].$$
In this modeling scenario, we assume the basis coefficients $\boldsymbol{\alpha}$ are such that $\alpha_{i} \sim \text{N}(\mu_{\alpha}, \sigma_{\alpha}^{2})$, $i = 1,\ldots,L$ and $\mathbf{c} \sim \text{N}(\mu_{c}\boldsymbol{1}, \sigma_{c}^{2} I),$  so that the priors on the basis coefficients are given by
\begin{align*}
\pi(\alpha_{i}|\mu_{\alpha}, \sigma_{\alpha}^{2}) &\propto (\sigma_{\alpha}^{2})^{-\frac{1}{2}}\text{exp} \left \{ -\frac{1}{2} (\alpha_{i} - \mu_{\alpha})^{2} / \sigma_{\alpha}^{2} \right \}
\end{align*}
and
\begin{align*}
\pi(\mathbf{c}|\mu_{c}, \sigma_{c}^{2} ) &= \left(2 \pi \sigma_{c}^{2}\right)^{-M/2} \exp\left\{-\frac{1}{2} \sum_{m=1}^{M} (c_{m} - \mu_{c})^{2} / \sigma_{c}^{2}\right\},
\end{align*}
for $i=1,\ldots,L$. The choice of hyper-parameters is similar to the previous model, i.e., we specify $\mu_{\alpha} = -10$, $\sigma_{\alpha}^{2} = 20$, $\mu_{c} = 0$ and $\sigma_{c}^{2} = 2.$ A similar behavior to the previous model was observed via extensive simulation studies, with a value of $\boldsymbol{\alpha} = (-30, -30, \ldots, -30)$ resulting in a high intensity of points, whereas, for $\boldsymbol{\alpha} = (10, 10, \ldots, 10)$ we obtain no points. Regarding the values of $\mathbf{c}$, we have noted that for $c = -2$ the model leads to strong attractions between events, whereas for a value $c = 2$ we noticed strong repulsion between events. We do not impose any restrictions on the interaction coefficients since the true interaction function is zero for IPPPs. Now, given that the point pattern arises from an IPPP, the estimated interaction function will tend to be either slightly above or below zero at different distances, i.e., even when we are estimating an IPPP model we do not anticipate a perfect estimate of $\mathbf{0}$ for $\mathbf{c}$, however, zero should always be in the credible set for each $c_{i}$ if the model is to be recovered correctly.

Based on the model distribution of (\ref{eqf}) and the aforementioned priors, the full conditional distribution of $\boldsymbol{\alpha}, \mathbf{c} | \cdot$ is given by
\begin{align*}
\pi \left(\boldsymbol{\alpha}, \mathbf{c} | \cdot \right) &= \frac{1}{Z_{n}(\boldsymbol{\alpha}, \mathbf{c})} \text{exp}\left[ -\left \{ \sum_{i=1}^{n} \boldsymbol{\phi}(\mathbf{x}_{i}) \right \}' \boldsymbol{\alpha} - \left \{ \sum_{\substack{i < j \\ i,j \ge 1}}^{n} \mathbf{b}(||\mathbf{x}_{i} - \mathbf{x}_{j}||) \right \}' \mathbf{c} \right] \\
&\times \prod_{i = 1}^{L} \left[(2\pi \sigma_{i}^{2})^{-\frac{1}{2}} \exp\left\{ -\frac{1}{2} (\alpha_{i} - \mu_{i})^{2} / \sigma_{i}^{2} \right\}\right] \\
&\times \left(2 \pi \sigma_{c}^{2}\right)^{-M/2} \exp\left\{-\frac{1}{2} \sum_{m=1}^{M} (c_{m} - \mu_{c})^{2} / \sigma_{c}^{2}\right\}.
\end{align*}
Based on this framework, the full conditional distribution for the $i$th global effect parameter is given by
\begin{align*}
\pi(\alpha_{i} | .) &\propto Z(\boldsymbol{\alpha}, \mathbf{c}) \text{exp}\left[ -\left\{ \sum_{l=1}^{n} \phi_{i}(\mathbf{x}_{l}) \right\}' \boldsymbol{\alpha} \right] \exp\left\{ -\frac{1}{2} (\alpha_{i} - \mu_{i})^{2} / \sigma_{i}^{2} \right\},
\end{align*}
and the full conditional distribution of the interaction parameters is given by
\begin{align*}
\pi(\mathbf{c} | .) &\propto Z(\boldsymbol{\alpha}, \mathbf{c}) \text{exp}\left[ - \left\{ \sum_{\substack{i < j \\ i,j \ge 1}}^{n} \mathbf{b}(||\mathbf{x}_{i} - \mathbf{x}_{j}||) \right\}' \mathbf{c} \right] \\
&\times \left(2 \pi \sigma_{c}^{2}\right)^{-M/2} \exp\left\{-\frac{1}{2} \sum_{m=1}^{M} (c_{m} - \mu_{c})^{2} / \sigma_{c}^{2}\right\}.
\end{align*}
For estimation of the model parameters, we make use of the Double Metropolis-Hastings algorithm discussed in Algorithm~\ref{algDoubleMH}, and we sample each $\alpha_{ij}$ individually.

\section{Applications}
\label{SectionApplications}


\subsection{Application to Water Strider Data}
\label{ApplicationsGibbs}
\begin{figure}[h]
	\begin{subfigure}[b]{0.475\textwidth}
		\centering
		 \includegraphics[angle=0,scale=0.45]{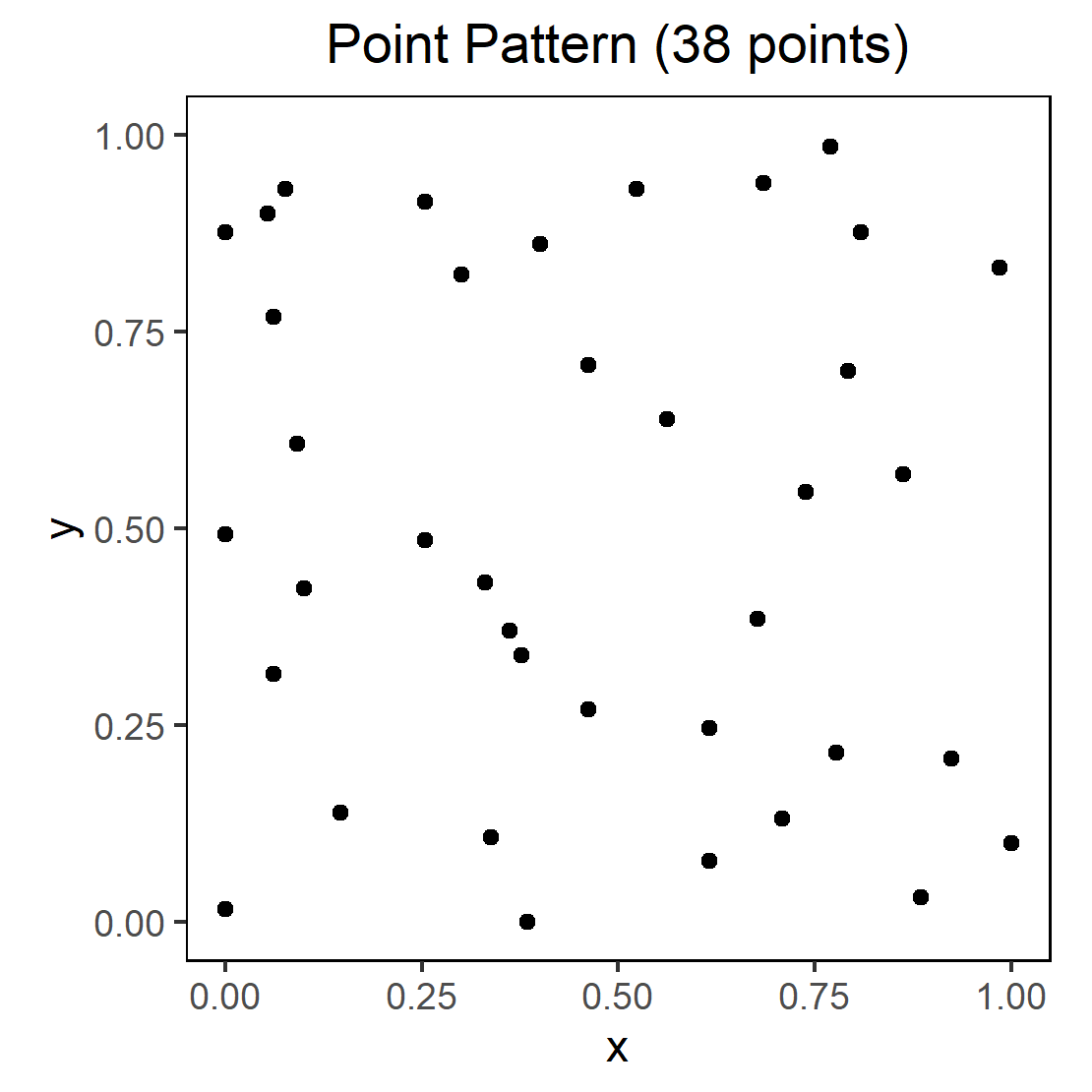}
		\caption{\baselineskip=10pt Locations of water striders from the first point pattern in the water striders dataset from the \textit{spatstat} R package.  Note that the point pattern was scaled to the window $[0,1]^{2}$.}
		\label{FigWaterstriders}
		\centering
	\end{subfigure}
	\hfill
	\begin{subfigure}[b]{0.475\textwidth}
		\centering
		\includegraphics[angle=0, scale=0.45]{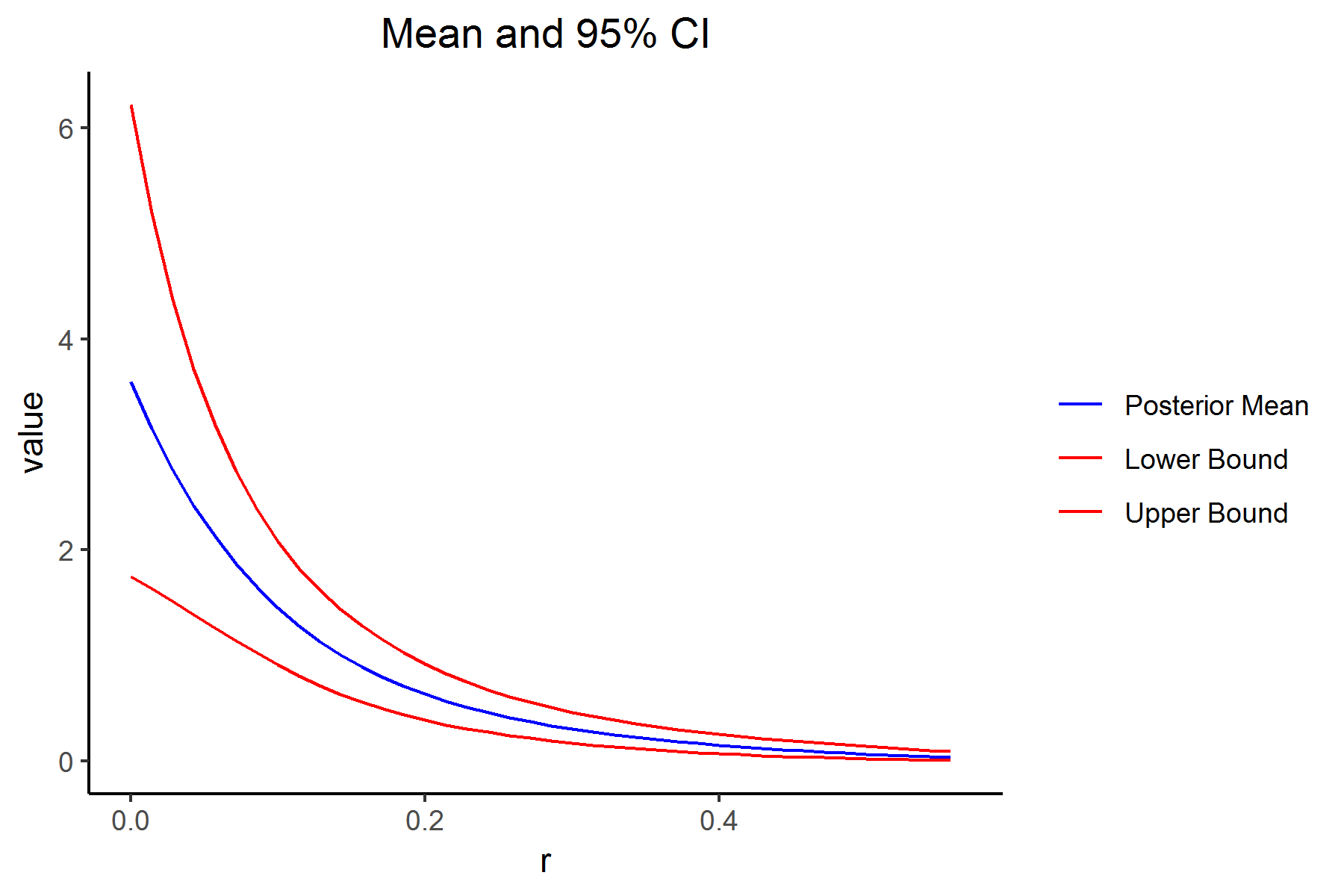}
		\caption{\baselineskip=10pt The interaction function for the waterstriders data.  Plots of the posterior mean interaction function (blue) and the 95\% Bayesian credible interval (red).}
		\label{FigWaterstriders_g_95_CI}
		\centering
	\end{subfigure}
	\vskip\baselineskip
	\begin{subfigure}[b]{0.475\textwidth}
		\centering
		\includegraphics[angle=0, scale=0.45]{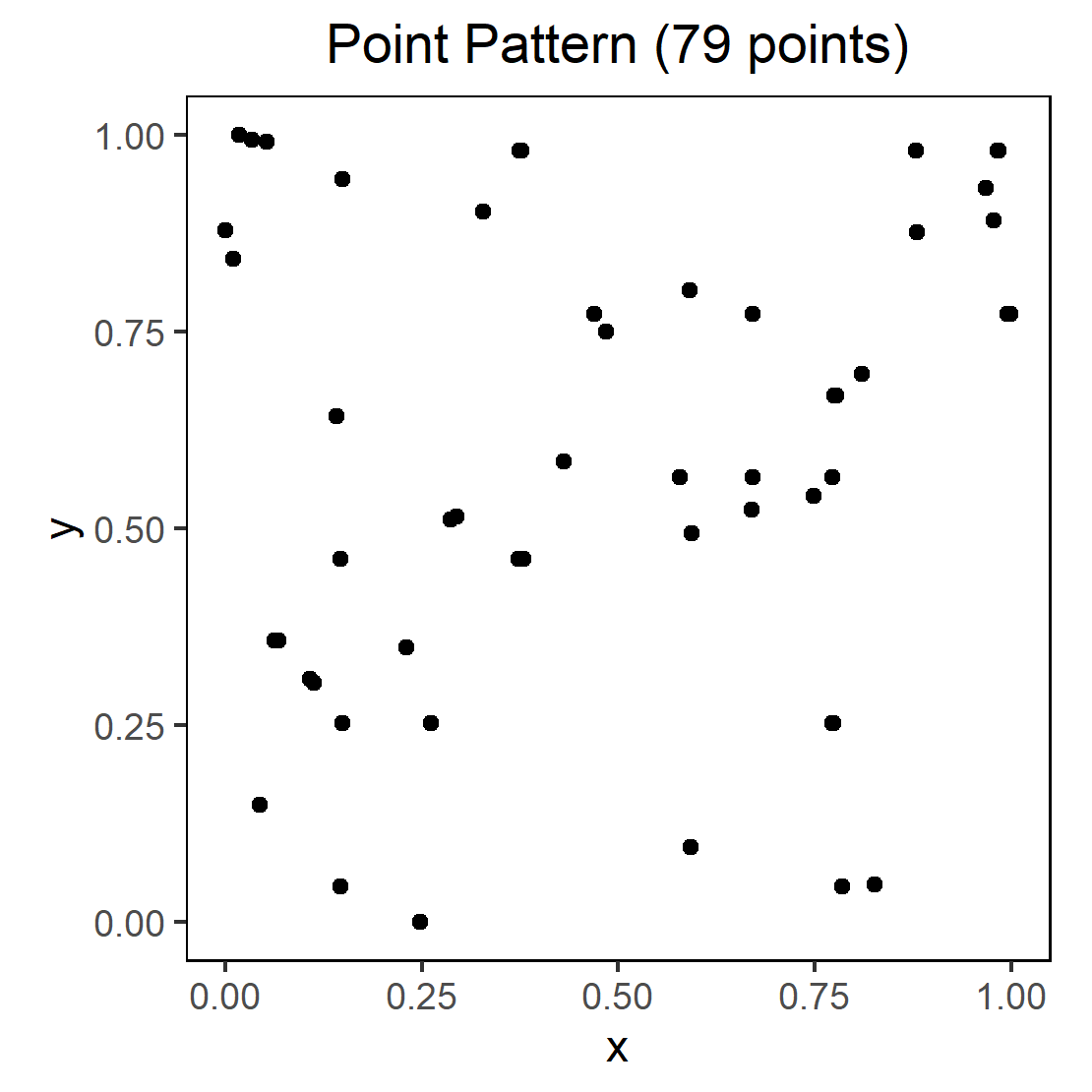}
		\caption{\baselineskip=10pt Locations of forest fires from 2005 to 2007 in the Castilla-La Mancha region of Spain.  The data are from the \texttt{clmfires} dataset from the \textit{spatstat} package in R, specifically events over $[100,200] \times [100,200].$ These events were shifted and scaled to $[0,1]^{2}$.}
		\label{Fig_clmfires}
		\centering
	\end{subfigure}
	\hfill
	\begin{subfigure}[b]{0.475\textwidth}
		\centering
		\includegraphics[angle=0, scale=0.45]{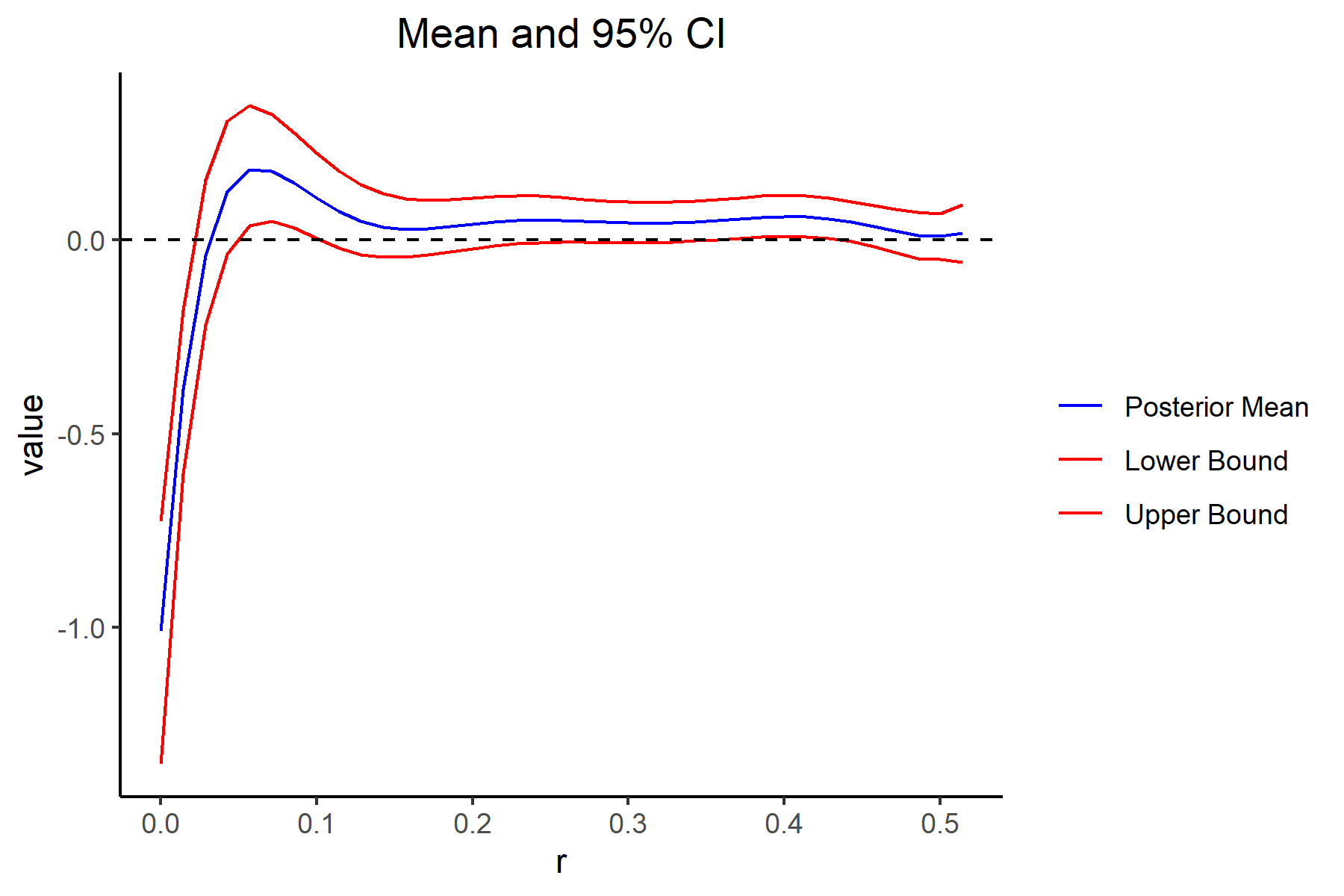}
		\caption{\baselineskip=10pt The interaction function of a subset of the \texttt{clmfires} dataset (See Section~\ref{IPPP_App_clmfires} for details).  Plot of the posterior mean interaction function (blue) and the 95\% Bayesian credible interval (red).}
		\label{Fig_clmfires_g_95_CI}
		\centering
	\end{subfigure}
	\caption{\baselineskip=10pt The water strider and forest fire data.}
	\label{Fig_ws_ff}
\end{figure}

We consider an application of our method to the water striders dataset from the \texttt{spatstat} {R} package \cite{baddeley2016spatial}.  The water striders dataset contains three different point patterns; we only consider the first one, and henceforth use the water striders dataset to refer to the first point pattern.  The events in this dataset are the locations of water striders in a pool, which features repulsion between the water striders as they are territorial.  First, we scale the locations of the water striders to be in $[0,1]^{2},$ and display the observed point pattern in Figure~\ref{FigWaterstriders}.  The points appear to occur in all areas of the window with no preference for one area over another, and there seems to be some distance between the points, as expected.

We fit our pairwise interaction model using the methodology previously described.  The sampler was run for $50,000$ iterations with the first $5,000$ used as burn-in.  A visual assessment of trace plots of the sample chains and the Gelman-Rubin test both indicated no lack of convergence.

A plot of the posterior mean of interaction functions and the $95\%$ CI of interaction functions is displayed in Figure~\ref{FigWaterstriders_g_95_CI}.  The interaction function has large values for small $r$ and then ``tapers" off, which is what we would expect due to the territorial nature of the water striders.  In fact, this interaction function looks similar to the one that was estimated for the Strauss model (\cite{strauss1975model}).  This is in line with the Strauss/hardcore interaction used in the literature to fit the water striders dataset \cite{penttinen1984modelling}. The posterior summary information is provided in Tables~\ref{TabWaterstriders_alpha}~and~\ref{TabWaterstriders_c}.

To assess the necessity of the interaction term, another model without interaction (one with $\mathbf{c} = \mathbf{0}$) was fit to the data, and the DIC for both models was calculated.  The model with interaction based on the posterior mean estimates of the parameters had a DIC value of $169.46,$ which was smaller than the model with zero interaction.  Therefore, our model indicates that there is interaction between points, which is what we would expect since the water striders are territorial.

Importantly, we note that our approach is able to obtain comparable results for a point pattern that was not simulated from our model.  From the global effect function we see that there is only a slight preference for points in one area of the window over another, whereas the interaction function indicates repulsion between the points.  Both of these observations are consistent with what would be expected for data featuring locations of water striders on a homogeneous surface.  Consequently, this example illustrates the flexibility of our model in capturing a diverse class of point patterns.

\begin{table}[ht]
	\centering
	\begin{tabular}{lccc}
		\hline
		Parameter & Mean & Lower Bound & Upper Bound \\
		\hline
		$\alpha_{1}$ & -9.65 & -12.95 & -6.64 \\
		$\alpha_{2}$ & -10.13 & -13.49 & -7.09 \\
		$\alpha_{3}$ & -12.79 & -16.76 & -9.22 \\
		$\alpha_{4}$ & -7.95 & -11.15 & -4.95 \\
		\hline
	\end{tabular}
	\caption{Water striders dataset.  Parameter summaries for the main effect function: means and 95\% credible
		intervals.}
	\label{TabWaterstriders_alpha}
\end{table}

\begin{table}[ht]
	\centering
	\begin{tabular}{lccc}
		\hline
		Parameter & Mean & Lower Bound & Upper Bound \\
		\hline
		$c_{1}$ & 3.60 & 1.75 & 6.22 \\
		$c_{2}$ & 1.50 & 0.61 & 2.75 \\
		$c_{3}$ & 0.81 & 0.35 & 1.48 \\
		$c_{4}$ & 0.51 & 0.22 & 0.93 \\
		$c_{5}$ & 0.33 & 0.14 & 0.60 \\
		$c_{6}$ & 0.22 & 0.09 & 0.41 \\
		$c_{7}$ & 0.15 & 0.05 & 0.30 \\
		$c_{8}$ & 0.10 & 0.03 & 0.22 \\
		$c_{9}$ & 0.06 & 0.01 & 0.16 \\
		$c_{10}$ & 0.03 & 0.00 & 0.09 \\
		\hline
	\end{tabular}
	\caption{Water striders dataset.  Parameter summaries for the interaction effect function: means and 95\% credible
		intervals.}
	\label{TabWaterstriders_c}
\end{table}


\subsection{Application to Forest Fire Data}
\label{IPPP_App_clmfires}
We consider an application of our method to the \texttt{clmfires} dataset from the \texttt{spatstat} {R} package \cite{baddeley2016spatial}.  The dataset features locations of forest fires from 1998 to 2007 in the Castilla-La Mancha region of Spain.  According to the package documentation, the precision of the events changed from 2003 to 2004, and as a result, we will only consider data from 2005 to 2007.  Since the data are over an irregular grid, we restrict our analysis to fires that occurred at locations that satisfy $x \in [100,200]$ and $y \in [100,200],$ which we shift and scale.  This results in a total of $79$ points, which can be seen in Figure~\ref{Fig_clmfires}.  From the plot, we see that there are fewer events in the bottom right of the window.  There are some points that are very close together, but there does not appear to be large clusters of points.  Therefore, by visual inspection, there does not appear to be any strong indication of interaction between the locations of the fires.

We fit our pairwise interaction model using the methodology described in Section~\ref{SectionSampling}.  The MCMC sampler was run for $100,000$ iterations with the first $10,000$ used as burn-in.  A visual assessment of trace plots of the sample chains and the Gelman-Rubin test both indicated no lack of convergence.

To assess the necessity of the interaction term, another model without interaction (one with $\mathbf{c} = \mathbf{0}$) was fit to the data, and the DIC for both models was calculated.  The model with interaction based on the posterior mean estimates of the parameters had a DIC that  was $23.09,$ which is smaller than the model with zero interaction.  Therefore, our model seems to indicate that there is some interaction between points. The $95\%$ Bayesian credible interval is presented in Figure~\ref{Fig_clmfires_g_95_CI}, further supports this initial conjecture. However, as the distance between events increases to about a distance of 0.13, we can clearly see that the events do not depend on each other anymore. This is not unusual, since forest fires tend to spread to nearby areas, forming clusters within short distances (within a few kilometers), while they are not affected by fires occurring very far away.

\begin{table}[ht]
	\centering
	\begin{tabular}{lccc}
		\hline
		Parameter & Mean & Lower Bound & Upper Bound \\
		\hline
		$\alpha_{1}$ & -5.54 & -7.35 & -3.85 \\
		$\alpha_{2}$ & -3.76 & -5.24 & -2.17 \\
		$\alpha_{3}$ & -5.79 & -7.34 & -4.30 \\
		$\alpha_{4}$ & -6.30 & -8.14 & -4.67 \\
		\hline
	\end{tabular}
	\caption{Forest fires dataset.  Parameter summaries for the main effect function: means and 95\% credible
		intervals.}
	\label{Tabclmfires_alpha}
\end{table}

\begin{table}[ht]
	\centering
	\begin{tabular}{lccc}
		\hline
		Parameter & Mean & Lower Bound & Upper Bound \\
		\hline
		$c_{1}$ & -1.01 & -1.35 & -0.72 \\
		$c_{2}$ & 2.29 & 1.53 & 3.15 \\
		$c_{3}$ & -2.11 & -3.44 & -1.00 \\
		$c_{4}$ & 0.84 & -0.48 & 2.19 \\
		$c_{5}$ & 0.62 & -0.89 & 2.25 \\
		$c_{6}$ & -0.83 & -2.04 & 0.68 \\
		$c_{7}$ & 0.39 & -0.77 & 1.21 \\
		$c_{8}$ & 0.20 & -0.51 & 0.88 \\
		$c_{9}$ & -0.11 & -0.50 & 0.25 \\
		$c_{10}$ & 0.04 & -0.09 & 0.17 \\
		\hline
	\end{tabular}
	\caption{Forest fires dataset.  Parameter summaries for the interaction effect function: means and 95\% credible
		intervals.}
	\label{Tabclmfires_c}
\end{table}

\section{Concluding Remarks}
\label{SectionConclusionAndFutureWork}


We proposed and illustrated a rich class of PP models via basis expansions of the main and interaction effects of a general Gibbs model.  We further provided a methodology that allows us to distinguish between a Gibbs process (dependent events) or a PPP (independent events). Model selection in order to determine the number of basis functions needed for a given dataset (such as stochastic search variable selection or the Bayesian Lasso) also constitute open problems in our proposed context and will be studied elsewhere.

\section*{Acknowledgments}
This article is released to inform interested parties of ongoing research and to encourage discussion. The views expressed on statistical issues are those of the authors and not those of the NSF. This research was partially supported by the U.S. National Science Foundation (NSF) under NSF grant NCSE-2215168.

\end{document}